\documentstyle[12pt,twoside]{article}
{\catcode `\@=11 \global\let\AddToReset=\@addtoreset}
\AddToReset{equation}{section}

\def\greaterthansquiggle{\raise.3ex\hbox{$>$\kern-.75em\lower1ex\hbox{$\sim$}}}
\def\lessthansquiggle{\raise.3ex\hbox{$<$\kern-.75em\lower1ex\hbox{$\sim$}}}
\newcommand{\beq}{\begin{equation}}
\newcommand{\eeq}{\end{equation}}
\newcommand{\beqa}{\begin{eqnarray}}
\newcommand{\eeqa}{\end{eqnarray}}
\newcommand{\beqan}{\begin{eqnarray*}}
\newcommand{\eeqan}{\end{eqnarray*}}
\newcommand{\ba}{\begin{array}}
\newcommand{\ea}{\end{array}}
\newcommand{\no}{\nonumber}

\newcommand{\ra}{\rightarrow}

\newcommand{\vp}{\varphi}

\newcommand{\dg}{\dagger}
\newcommand{\wt}{\widetilde}
\newcommand{\wh}{\widehat}

\newcommand{\A}{{\cal A}}

\newcommand{\D}{{\cal D}}

\newcommand{\F}{{\cal F}}
\newcommand{\G}{{\cal G}}
\newcommand{\Ha}{{\cal H}}

\newcommand{\M}{{\cal M}}

\newcommand{\st}{\stackrel}
\newcommand{\dfrac}{\displaystyle \frac}

\def\nz{\ifmmode {I\hskip -3pt N} \else {\hbox {$I\hskip -3pt N$}}\fi}
\def\zz{\ifmmode {Z\hskip -4.8pt Z} \else
       {\hbox {$Z\hskip -4.8pt Z$}}\fi}
\def\qz{\ifmmode {Q\hskip -5.0pt\vrule height6.0pt depth 0pt
       \hskip 6pt} \else {\hbox
       {$Q\hskip -5.0pt\vrule height6.0pt depth 0pt\hskip 6pt$}}\fi}
\def\rz{\ifmmode {I\hskip -3pt R} \else {\hbox {$I\hskip -3pt R$}}\fi}
\def\cz{\ifmmode {C\hskip -4.8pt\vrule height5.8pt\hskip 6.3pt} \else
       {\hbox {$C\hskip -4.8pt\vrule height5.8pt\hskip 6.3pt$}}\fi}
\def\au{{\setbox0=\hbox{\lower1.36775ex%
\hbox{''}\kern-.05em}\dp0=.36775ex\hskip0pt\box0}}
\def\ao{{}\kern-.10em\hbox{``}}

\normalsize
\begin{document}
\bibliographystyle{plain}

\begin{titlepage}
\begin{flushright}
UWThPh-1998-4
\end{flushright}
\vspace{2cm}
\begin{center}
{\Large \bf Generalized Stochastic Quantization \\[5pt]
of Yang-Mills Theory} \\[40pt]
Helmuth H\"uffel* and Gerald Kelnhofer** \\
Institut f\"ur Theoretische Physik \\
Universit\"at Wien \\
Boltzmanngasse 5, A-1090 Vienna, Austria
\vfill

{\bf Abstract}
\end{center}

We perform the stochastic quantization of Yang--Mills theory in 
configuration space and derive the Faddeev-Popov path integral density.
Based on a 
generalization of the stochastic gauge fixing scheme and its geometrical 
interpretation this result is 
obtained as the exact equilibrium solution of the associated
Fokker--Planck equation. Included in our discussion is the  
precise range of validity of our approach.

\vfill
\begin{enumerate}
\item[*)] Email: helmuth.hueffel@univie.ac.at
\item[**)] Supported by "Fonds zur F\"orderung der wissenschaftlichen 
Forschung in \"Oster\-reich",  project P10509-NAW
\end{enumerate}
\end{titlepage}

\section{Introduction}

The stochastic quantization method of Parisi and Wu  \cite{Parisi+Wu}
was introduced 
1981 as a new method for quantizing field theories. It is based on 
concepts of nonequilibrium statistical mechanics and provides novel and
alternative 
insights into quantum field theory, see refs.
\cite{Damgaard+Huffel,Namiki} for 
comprehensive reviews 
and referencing. One of the most interesting aspects of this new
quantization 
scheme lies in 
its rather unconventional treatment of gauge field theories, in specific
of
Yang-Mills theories. We do not intend to 
review here the basic  
facts, benefits or problems of the stochastic quantization scheme of 
gauge field theories (see, however, \cite{annals}) but just recall that 
originally it was 
formulated  by Parisi and Wu  without 
the introduction of gauge fixing terms and the usual 
Faddeev-Popov ghost 
fields; later on a modified approach named stochastic gauge fixing
was given by Zwanziger \cite{Zwanziger81} where again no Faddeev-Popov 
ghost fields where introduced. 
Our focus is based 
on extending a previously 
introduced generalization \cite{StMargh,Physlett,annals} of Zwanziger's 
stochastic gauge fixing 
scheme. We so far 
studied the 
helix model \cite{deWit,Kuchar,Friedberg et al.,Fujikawa} 
which is an abelian gauge theory coupled to 
bosonic matter fields in $0+1$ 
dimensions which does not suffer from a Gribov ambiguity \cite{Gribov}.
By this 
generalized stochastic gauge fixing scheme it was possible to derive 
a non perturbative proof of the equivalence between  
the conventional path integral formulation of this model and the 
equilibrium limit of the corresponding stochastic correlation functions. 
The method mainly is based on the possibility to introduce adapted 
coordinates which means to separate the original 
gauge fields into gauge independent and gauge dependent degrees of 
freedom. 

In the present article we straightforwardly generalize our formalism to
the 
nonabelian Yang--Mills theory. In comparison with the helix model,
however, the 
geometrical structure of Yang--Mills theory obstructs a global
separation
of the field 
variables as mentioned above due 
to the well known Gribov ambiguity. Therefore we have to restrict our 
investigation to sufficiently small regions in the space of Yang--Mills 
fields. 

The main difficulty in the previous investigations of the stochastic 
quantization of Yang--Mills theory for  deriving a conventional field
theory 
path integral density was to solve the Fokker--Planck equation 
in the equilibrium limit. 
In the original Parisi--Wu approach this equilibrium limit could not
even be 
attained due to unbounded diffusions of the gauge modes.
Zwanziger \cite{Zwanziger81,Baulieu+Zwanziger,Zwanziger86}
 suggested to introduce a specific additional  nonholonomic
stochastic force term to 
suppress these gauge modes 
yet keeping the expectation values of 
gauge invariant observables unchanged. The approach to equilibrium and 
the discussion of the 
conditions of applicability to the nonperturbative regime, 
however, do not seem to have been fully completed.

Our analysis is distinguished by the above approaches 
by exploiting a more general freedom to modify 
both the drift term and the diffusion term of the stochastic process
again leaving all expectation values 
of gauge invariant variables unchanged. Due to this additional structure 
of modification the equilibrium limit could be obtained immediately
using 
the fluctuation dissipation theorem proving equivalence with the well 
known Faddeev-Popov path integral density. In deriving this result the 
gauge degrees of freedom were fully under control, no infinite gauge 
group volumes arose. However, this equivalence proof has been performed 
only for those gauge field configurations satisfying a unique gauge 
fixing condition leaving the option for further investigations
concerning the 
Gribov issue.

In section 2 the geometrical setting for Yang--Mills theory is 
introduced. The relevant objects are identified and the non trivial 
bundle structure of the space of gauge potentials is outlined. 

Section 3 offers a brief review on the stochastic gauge fixing scheme  
issued by Zwanziger.

A new generalized stochastic 
gauge fixing method for Yang--Mills theory is presented in section 4. 
We exploit the most general form of 
the Langevin equation such that the expectation values of 
gauge invariant observables remain unchanged.The 
adapted coordinates, the corresponding vielbeins and metrics are 
defined.

The geometrical structure of the generalized stochastic gauge fixing is 
revealed in depth in section 5. Due to the extension of the stochastic 
process a new metric is induced with respect to which the 
space spanned by the gauge invariant fields becomes orthogonal to the
gauge 
orbits. The relation between this metric and several horizontal bundles 
in the space of gauge fields is elucidated. 
 
Section 6 is devoted to the derivation of the path integral density as
an 
equilibrium solution of the Fokker--Plank equation. Thereby the 
equivalence with the Faddeev--Popov approach is proved. 

Finally an outlook is presented in section 7.
\section{The geometrical setting of Yang-Mills theory}
In this section we present the major geometrical structures of pure
Yang-Mills theory. We collect in a somewhat formal style all the
necessary ingredients which are needed later on for a compact and 
transparent formulation of the stochastic quantization scheme of
Yang-Mills theory.

Let $P(M,G)$ be a principal fiber bundle with structure group $G$ over
the compact Euclidean space time $M$. Let $\A$ denote the space of all
irreducible connections  on $P$ and let $\G$ denote the gauge group, 
which is given by all vertical automorphisms on $P$ reduced by the
centre
of $G$. Then $\G$ acts freely on $\A$ and defines a principal
$\G$-fibration
$\A \st{\pi}{\longrightarrow} \A/\G =: \M$ over the space
$\M$ of all inequivalent gauge potentials
with projection $\pi$ \cite{Singer,Nara,Mitter}. 
A Riemannian structure on $\A$ can be 
introduced as follows: Let $g$ denote the Lie algebra of $G$ and
consider
the adjoint bundle ad~$P = P \times_{\rm ad} g$ which is associated to
the
principal bundle $P$ via the adjoint action of $G$ on $g$. Choosing the
natural Killing form on $g$ an inner product can be defined on the space
$\Omega^q(M,\mbox{ad }P)$ of ad~$P$-valued $q$-forms on $M$ by
\beq
\langle \vp,\vp'\rangle_{(q)} = \int_M \mbox{tr}(\vp \wedge * \vp')
\eeq
where $*$ is the Hodge operation with respect to the given metric on $M$ 
and $tr$ denotes the trace on $g$.
Locally a form in $\Omega^q(M,\mbox{ad }P)$ is just a $g$-valued
$q$-form
on $M$.

Since $\A$ is an affine space modelled  on $\Omega^1(M,\mbox{ad }P)$ the
tangent bundle of $\A$ is given by $T \A = \A \times \Omega^1(M,
\mbox{ad }P)$. Hence one can define a Riemannian structure $h$ on $\A$
by
\beq
h_A : T_A \A \times T_A \A \ra R, \qquad
h_A(\tau^1,\tau^2) := \langle \tau^1,\tau^2\rangle_{(1)}  \qquad \tau^1, 
\tau^2 \in \Omega^1(M,\mbox{ad }P).
\eeq
The space $\Omega^0(M,\mbox{ad }P)$ can be identified with the
Lie algebra $Lie~\G$ of the gauge group $\G$ and a natural
inner product on $Lie~\G$ is given by $\langle \; ,\;\rangle_{(0)}$.

Due to the Gribov ambiguity \cite{Gribov} the principal $\G$-bundle $\A
\ra \M$
is not globally trivializable. In order to define a local section, we
choose a fixed background connection $A_0 \in \A$ and consider a
sufficient small neighbourhood $U(A_0)$ of $\pi(A_0)$ in $\M$.
Then the subspace
\beq
\Sigma = \{ B \in \pi^{-1}(U(A_0))/D_{A_0}^*(B - A_0) = 0\}
\eeq
defines a local section of $\A \ra \M$ \cite{Atiyah,Mitter}. 
Here $D_{A_0}^*$ is the
adjoint operator of the covariant derivative $D_{A_0}$ with respect
to $A_0$. A tangent vector $\zeta_B \in T_B \Sigma$ is uniquely
characterized by
the property $D_{A_0}^* \zeta_B = 0$.

Notice that in the zero instanton sector (for $M = S^4$)
this background field $A_0$ can be set to zero, yielding the familiar 
covariant gauge condition
\beq
\partial_\mu B^\mu = 0.
\eeq
\section{Stochastic gauge fixing}

We start with the Parisi--Wu approach for the stochastic quantization
of the Yang--Mills theory in terms of the Langevin equation
\beq
dA = - \frac{\delta S}{\delta A} ds + dW.
\eeq
Here $S$ denotes the Yang--Mills action without gauge symmetry breaking
terms and without accompanying ghost field terms
\beq
S = \frac{1}{2} \langle F,F\rangle_{(2)},
\eeq
where $F$ denotes the curvature of $A$,
$s$ denotes the extra time coordinate (``stochastic time'' coordinate)
with respect to which the stochastic process is evolving, $dW$ is the
increment of a Wiener process (for a detailed presentation see e.g. 
\cite{Arnold}).

We now discuss Zwanziger's modified formulation \cite{Zwanziger81} of 
the Parisi-Wu scheme: The stochastic gauge fixing procedure 
consists in adding an additional drift force to the Langevin 
equation (3.1) which acts tangentially to the gauge orbits. This 
additional term generally can be expressed by the gauge 
generator $D_A$ and an arbitrary function $\alpha$
 so that the modified Langevin equation reads as follows
 \beq
dA = \left[ -\frac{\delta S}{\delta A} + D_A \alpha 
\right] ds
+ dW.
\eeq

The expectation values of gauge invariant observables remain 
unchanged for any choice of the function $\alpha$ (see below for the 
explicit demonstration contained in the discussion of our 
generalized stochastic gauge fixing procedure).
For specific choices of the -- in principle -- arbitrary 
function 
$\alpha$ the gauge modes' diffusion is damped along the gauge 
orbits. As a consequence the Fokker-Planck density can be 
normalized; we remind that this situation is in contrast to the Parisi-
Wu approach, where for expectation values of gauge variant 
observables no equilibrium values could be attained.

In contrast to the approach of \cite{Zwanziger81} where no equilibrium 
distribution of the Fokker--Planck equation could be derived as well as 
in contrast to \cite{Baulieu+Zwanziger} where the full Fokker--Planck 
operator 
\beq
L = \frac{\delta }{\delta A}\left[  \frac{\delta S}{\delta A} - D_A 
\alpha + \frac{\delta }{\delta A} 
\right] 
\eeq       
was needed to obtain an equilibrium distribution we present a 
quite different strategy:

As the Fokker-Planck operator factorizes into first order differential 
operators  the question arises whether it is possible to derive the 
equilibrium distribution directly by solving a simpler first order 
problem. However, for this to be possible a necessary integrability 
condition imposed on the drift term $\dfrac{\delta S}{\delta A} - D_A 
\alpha$ has to be fulfilled. It is 
well known that for the Yang--Mills case this is violated.   

In the following we want to clarify the relationship of this 
integrability condition and 
the underlying geometrical structure of the space of gauge potentials.

We remind that any bundle metric on a principal fiber
bundle which is invariant under the corresponding group action gives
rise
to a natural connection whose horizontal subbundle is orthogonal to the
corresponding group. The natural connection induced by 
$\langle \; ,\;\rangle_{(1)}$ in Yang-Mills theory is given by the 
following $Lie\G$ valued one form
\beq
\gamma = \Delta_A^{-1} D_A^*.
\eeq
The projection ${\bf P}$ onto the corresponding horizontal subbundle is
given by
\beq
{\bf P} = {\bf 1} - D_A \gamma.
\eeq
The curvature $\Omega$ of $\gamma$, $\Omega = \delta_\A \gamma + 
\frac{1}{2} [\gamma,\gamma]$, where $\delta_\A$ denotes the exterior 
derivative on $\A$, however, does not vanish \cite{Singer} so that there
does
not exist (even locally) a manifold whose tangent bundle is isomorphic
to this horizontal subbundle. Moreover this also implies that any vector
field along the gauge group cannot be written as a gradient of a
function 
with respect to the metric $\langle \; , \;\rangle_{(1)}$. 

To verify
this explicitly let us assume that the one form $\langle
D_A\alpha(A),\cdot
\rangle$, where $\alpha(A)$ is any $Lie~\G$-valued function on $\A$ is
the
differential of a function $f$ on $\A$, i.e.
\beq 
\langle D_A \alpha(A),\cdot\rangle_{(1)} = \delta_\A f.
\eeq
For two vector fields $\tau^1$, $\tau^2$ in $T\A$ being 
horizontal with
respect to $\gamma$ (i.e. $\gamma(\tau^1) = \gamma(\tau^2) = 0$)
we have
\beq
\Omega(\tau^1,\tau^2) = - \gamma([\tau^1,\tau^2]),
\eeq
so that the one 
form on the left hand side of (3.7) gives on the vector field commutator
$[\tau^1,\tau^2]$
\beq
\langle D_A \alpha(A),[\tau^1,\tau^2]\rangle_{(1)} =
\langle D_A \alpha(A),({\bf 1} - {\bf P}_A)[\tau^1,\tau^2]\rangle_{(1)}
= - \langle \alpha(A),\Delta_A \Omega(\tau^1,\tau^2) \rangle \neq 0.
\eeq
But $\delta_\A f[\tau^1,\tau^2] = 0$ since $\delta_\A f(\tau^1) =
\delta_\A f(\tau^2) = 0$ hence giving a contradiction. However it
should be remarked that the vanishing of the curvature is only a
necessary condition.

It is our intention to modify the stochastic process (3.3) for the 
Yang--Mills theory in such a 
way that the factorization of the modified Fokker-Planck operator indeed
allows 
the determination of the equilibrium distribution as a solution of a
first 
order differential equation in a consistent manner.

\section{Generalized stochastic gauge fixing}

In this section we apply our recently introduced
\cite{StMargh,Physlett,annals}
generalization of the
stochastic gauge fixing procedure of Zwanziger to the Yang--Mills
theory. It is  
advantageous to avoid the complicated nonabelian dynamics of 
the Yang-Mills fields by transforming them into a set of adapted 
coordinates \cite{Boulw,Zwanziger86}. This means to separate the
original 
gauge fields into gauge independent and gauge dependent degrees of 
freedom. However, this is only locally possible due to the non
triviality 
of the bundle $\A \ra \M$ 
so that we are forced to consider the trivializable
bundle $\pi^{-1}(U(A_0)) \ra U(A_0)$. 
In concrete, our analysis will be performed on the isomorphic trivial
principal $\G$-bundle, $\Sigma \times \G \ra \Sigma$, where the
isomorphism
is given by the map
\beq
\chi : \Sigma \times \G \ra \pi^{-1}(U(A_0)), \qquad
\chi(B,g) := B^g
\eeq
with $B \in \Sigma$, $g \in \G$ and $B^g$ denoting the nonabelian gauge 
transformation of $B$ by $g$
\beq
B^g = g^{-1}Bg + g^{-1}dg.
\eeq
Evidently the inverse map
$\chi^{-1}$ is given by the expression
\beq
\chi^{-1} : \pi^{-1}(U(A_0)) \ra \Sigma \times \G, \qquad
\chi^{-1}(A) := (A^{\omega(A)^{-1}}, \omega(A)),
\eeq
where $\omega : \pi^{-1}(U(A_0)) \ra \G$ is uniquely defined by the
requirement that $A^{\omega(A)^{-1}} \in \Sigma$, i.e.
\beq
D_{A_0}^*(A^{\omega(A)^{-1}} - A_0) = 0.
\eeq
Although an explicit expression for
$\omega$ can be given only in terms of a perturbative expansion 
\cite{Boulw}, 
it is nevertheless easy to derive in closed form its differential, 
which is necessary to calculate the corresponding
vielbeins. To do this, we begin by calculating the differential
$T\chi$ of $\chi$. $T\chi$ provides an isomorphism 
$T\chi : T(\Sigma \times \G) \ra T \pi^{-1}(U(A_0))$ given by
\beq
T\chi(\zeta_B,Y_g) = \mbox{ad}(g^{-1})(\zeta_B + D_B R_g(Y_g))
\eeq
where $\zeta_B \in T_B \Sigma \subset \Omega^1(M,\mbox{ad }P)$, $Y_g$ is
a 
tangent vector on the gauge group $\G$ in point $g$, ad denotes the
adjoint action of $\G$ on ~$Lie\G$, and $R_g$ denotes the right
logarithmic 
derivative of the
identity on $\G$ (i.e. $R_g$ is that invertible operator on $T\G$ which
transports a tangent vector in $T_g \G$ back to the identity by the
differential of right multiplication).

From (4.1) the vielbeins $e$ corresponding to the change of coordinates
$(B,g) \ra A$ are given by
\beq
e = \left(  e_\Sigma, e_\G  \right) = \mbox{ad}(g^{-1})
\left( {\bf P}^\Sigma, D_BR_g  \right).
\eeq
Here ${\bf P}^\Sigma = {\bf 1} - D_{A_0} \Delta_{A_0}^{-1} D_{A_0}^*$
denotes the projector onto the subspace $T_B\Sigma$ and
$\Delta_{A_0}^{-1}$
is the inverse of the covariant Laplacian $\Delta_{A_0} = D_{A_0}^*
D_{A_0}$.

Now it is an easy task to verify that the following map is the inverse
of $T\chi$, hence giving the tangent map of $\chi^{-1}$, namely
$$
T \chi^{-1} : T(\pi^{-1}(U(A_0))) \ra T(\Sigma \times \G),
$$
\beq
T\chi^{-1}(\tau_A) = ({\bf P}^\Sigma({\bf 1} - D_B \F_B^{-1}
D_{A_0}^*) \mbox{ad}(g) \tau_A, R_g^{-1} \F_B^{-1}
D_{A_0}^* \mbox{ad}(g) \tau_A), 
\eeq
where $\tau_A \in T_A(\pi^{-1}(U(A_0)))$. Here $B=A^{\omega(A)^{-1}}$, 
$g=\omega(A)$ and 
\beq
 \F_B: \Omega^0(M,\mbox{ad }P) \ra \Omega^0(M,
\mbox{ad }P), \qquad \F_B = D_{A_0}^* D_B
\eeq
denotes the Faddeev--Popov operator. Since for sufficiently small
$U(A_0)$ the coordinate transformation onto the adapted coordinates is 
regular the Faddeev--Popov operator $\F_B$ is invertible, and 
$\Sigma$ thus is completely contained within one Gribov horizon.
Notice that the Faddeev--Popov operator 
 is self-adjoint for all
$B \in \Sigma$. 

From (4.3) the vielbeins $E$ corresponding to the change of coordinates
$A \ra (B,g)$ are given by
\beq
E = \left( \ba{c} E^\Sigma \\ E^\G \ea \right) =
\left( \ba{c} {\bf P}^\Sigma({\bf 1} - D_B \F_B^{-1} D_{A_0}^*)
\mbox{ad}(g) \\
R_g^{-1} \F_B^{-1} D_{A_0}^* \mbox{ad}(g) \ea \right) .
\eeq
Before we defined a Riemannian structure on $\A$ by the inner product
$\langle \; ,\;\rangle_{(1)}$. However, in the adapted coordinates
$(B,\omega)$ this metric $G$ is given as follows (pullback of $h$ by
$\chi$), 
\beqa
\lefteqn{G_{(B,g)} ((\zeta_B^1,Y_g^1),(\zeta_B^2,Y_g^2))
= } \no  \\
&=& \langle \zeta_B^1,\zeta_B^2\rangle_{(1)} +
\langle \zeta_B^1,D_B R_g(Y_g^2)\rangle_{(1)} +
\langle D_B R_g(Y_g^1),\zeta_B^2\rangle_{(1)} +
\langle R_g(Y_g^1),\Delta_B R_g(Y_g^1)\rangle_{(1)} \no
\\
\eeqa
where $\zeta_B^1,\zeta_B^2 \in T_B \Sigma$ and $Y_g^1,Y_g^2 \in T_g \G$.

Formally, the metric $G$ can be written in matrix form
\beq
G = e^*e = \left( \ba{cc} {\bf P}^\Sigma & 
{\bf P}^\Sigma \cdot D_B \cdot R_g \\[9pt]
R_g^* \cdot D_B^* \cdot {\bf P}^\Sigma &
R_g^* \cdot \Delta_B \cdot R_g \ea \right)
\eeq
where $R_g^*$ is the adjoint operation of $R_g$ with respect to the
inner
product $\langle \; ,\;\rangle_{(0)}$ on $Lie~\G$. We also mention the
inverse metric $G^{-1}$
\beqa
G^{-1} &=&
\left( \ba{cc}
(G^{-1})^{\Sigma \Sigma} &
(G^{-1})^{\Sigma \G} \\[6pt]
(G^{-1})^{\G \Sigma} &
(G^{-1})^{\G \G} \ea \right) = 
EE^* = \no \\[9pt]
&=&
\left( \ba{cc}
{\bf P}^\Sigma - {\bf P}^\Sigma D_B \F_B^{-1} \Delta_{A_0} \F_B^{-1}
D_B^*
{\bf P}^\Sigma &
{\bf P}^\Sigma D_B \F_B^{-1} \Delta_{A_0} \F_B^{-1} R_g^{*-1} \\[6pt]
R_g^{-1} \F_B^{-1} \Delta_{A_0} \F_B^{-1} D_B^* &
- R_g^{-1} \F_B^{-1} \Delta_{A_0} \F_B^{-1} R_g^{*-1} \ea \right).
\eeqa
The determinant of $G$ is then given by 
\beq
\det G = \det (R_g^* R_g)(\det \F_B)^2 (\det \Delta_{A_0})^{-1}
\eeq
where $\sqrt{\det (R_g^* R_g)}$ can be identified with the volume
density 
on $\G$, associated to the (right) invariant metric $R^*R$ on $\G$.

In the following we transform the Parisi--Wu Langevin equation (3.1)
into 
the adapted coordinates $\Psi = \left( \ba{c} B \\ g \ea \right)$. As 
this transformation is not globally possible the region of 
definition of (3.1) has to be restricted to
$\pi^{-1}(U(A_0))$. Making use of the Ito stochastic calculus
\cite{Arnold,annals} the above Langevin equation now reads 
\beq
d\Psi = \left( - G^{-1} \frac{\delta S}{\delta \Psi} + 
\frac{1}{\sqrt{\det G}} \frac{\delta(G^{-1} \sqrt{\det G})}{\delta \Psi}
\right)ds + E dW
\eeq
where the vielbein $E$, the metric $G$, its inverse and its determinant
were introduced in the previous section.

The generalized stochastic quantization procedure amounts -- as a direct
consequence of our previous investigations \cite{annals} on the abelian
helix model
 -- to consider the modified Langevin equation
\beq
d\Psi = \left( -G^{-1} \frac{\delta S}{\delta \Psi} + 
\frac{1}{\sqrt{\det G}} \frac{\delta(G^{-1}\sqrt{\det G})}{\delta\Psi}
 + E D_A \alpha \right) ds + E({\bf 1} + D_A \beta)dW,  
\eeq
where $A=B^g$. Here $\alpha : \pi^{-1}(U(A_0)) \ra Lie \G$ 
and the $Lie \G$ valued one form 
$\beta \in \Omega^1(\pi^{-1}(U(A_0)),Lie\G)$
are \`a priori arbitrary and will be fixed later on. 

The above Langevin 
equation is the most general 
Langevin equation for Yang--Mills theory which leads to the same 
expectation values of gauge invariant variables as the original 
Parisi--Wu equation (3.1)  written in adapted coordinates:

Let us remind that
the stochastic time evolution of expectation values of observables is 
described by the adjoint Fokker--Planck operator
$L^\dg$ which corresponding to (4.15) is given by
\beq
L^\dg =
\left[ -\frac{\delta S}{\delta \Psi} +
\frac{1}{\sqrt{\det G}} \; \frac{\delta \sqrt{\det G}}{\delta \Psi}
+ \frac{\delta}{\delta \Psi}\right]
G^{-1} \frac{\delta}{\delta \Psi}
+L^+_{\rm extra}
\eeq
We introduce
\beq
\wt E = E({\bf 1} + D_A \beta), \qquad \wt G^{-1} = \wt E \wt E^*, 
\eeq
with $A=B^g$ and have
\beq
L^+_{\rm extra} = (E D_A \alpha)^* \frac{\delta}{\delta \Psi} 
+(\wt G^{-1} - G^{-1}) \frac{\delta^2}{\delta \Psi \delta \Psi}
\eeq
where again $A=B^g$
and where the second term in (4.18) reads explicitly 
\beqa
(\wt G^{-1} - G^{-1}) \frac{\delta^2}{\delta \Psi \delta \Psi} &=&
(\wt G^{-1} - G^{-1})^{\Sigma\Sigma} \frac{\delta^2}{\delta B \delta B} 
+(\wt G^{-1} - G^{-1})^{\Sigma\G} \frac{\delta^2}{\delta B \delta 
g} \no \\[9pt]
&&+(\wt G^{-1} - G^{-1})^{\G\Sigma} \frac{\delta^2}{\delta g \delta B} 
+(\wt G^{-1} - G^{-1})^{\G\G} \frac{\delta^2}{\delta g \delta g}. 
\eeqa
Our proof consists in showing that the $\alpha,\beta$ dependent
extra term $L^+_{\rm extra}$ of $L^+$ annihilates on gauge invariant
observables. 
Indeed we obtain from (4.9)
\beq
(E D_A \alpha)^{*\Sigma} = 0, 
\eeq
where $A=B^g$. Furthermore we have
\beq
(\wt G^{-1} - G^{-1})^{\Sigma\Sigma} = 0
\eeq
so that the action of $L^+_{\rm extra}$ on gauge invariant observables,
which are 
purely functions $f(B)$ when written in terms of adapted coordinates,  
identically vanishes
\beq
L^+_{\rm extra} \; f(B) = 0.
\eeq

Alternatively we observe that due to (4.20) the $\alpha$ and
$\beta$ dependent terms in the modified Langevin equation (4.15) drop
out after projecting on the gauge invariant subspace $\Sigma$ described 
by the coordinate $B$
\beq
dB = \left[ -(G^{-1})^{\Sigma \Sigma} \frac{\delta S}{\delta B} + 
\frac{1}{\sqrt{\det G}} \frac{\delta((G^{-1})^{\Sigma \Sigma}\sqrt{\det
G})}
{\delta B} \right]ds + E^\Sigma dW.
\eeq
In deriving the above Langevin equation we have used the fact 
that the divergence of the 
generator of right group transformations corresponding to the invariant 
group measure induced by the metric $R_g^* R_g$ vanishes, i.e.
\beq
\frac{1}{\sqrt{\det (R_g^* R_g)}} 
\frac{\delta(\sqrt{\det (R_g^* R_g)} R_g^{*-1})}{\delta g} = 0
\eeq

We close by transforming back the Langevin equation (4.15) into the
original coordinates $A$. Invoking the Ito stochastic calculus once more
again we have
\beq
dA = \left[ -\frac{\delta S}{\delta A} + D_A \alpha + 
\frac{\delta^2 A}{\delta \Psi \delta \Psi} (\wt G^{-1} - G^{-1}) 
\right] ds
+ ({\bf 1} + D_A \beta)dW.
\eeq
In the above equation it is understood to take $B=A^{\omega (A)^{-1}}$ 
and $g= \omega (A)$.
Let us remind that the above Langevin equation  is valid only in the 
restricted domain $\pi^{-1}(U(A_0))$. 

\section{On the geometrical interpretation of generalized stochastic
gauge fixing}

As a consequence of our generalized stochastic gauge fixing procedure
not only Zwan\-ziger's original term $D_A \alpha$ is appearing in the
Langevin equation (4.25) for the Yang--Mills field $A$, but also an
additional $\beta$-dependent drift term as well as a specific
modification
of the Wiener increment, described by the operator
\beq 
\wh e = {\bf 1} + 
D_A \beta .
\eeq
We regard $\wh e$ as a $T \pi^{-1}(U(A_0))$-valued one form 
on $\pi^{-1}(U(A_0))$ by
setting $\wh e(\tau) = \tau + D_A \beta(\tau)$ for all tangent
vectors $\tau \in T_A \pi^{-1}(U(A_0))$.
The idea is to view
$\wh e$ as a vielbein giving rise to the inverse of a yet
not specified metric $\wh g$ on the space $\pi^{-1}(U(A_0))$.
We note that the inverse vielbein $\wh e^{-1}$
is given by the $T\pi^{-1}(U(A_0))$-valued one form on
$\pi^{-1}(U(A_0))$
\beq
\wh e^{-1} = {\bf 1} - D_A({\bf 1} + \beta D_A)^{-1} \beta.
\eeq
provided the operator ${\bf 1} + \beta D_A:Lie \G \ra
Lie \G$ is invertible for all $A \in \pi^{-1}(U(A_0))$.
Hence the metric $\wh g = \wh e^{-1*} \wh e^{-1}$ is given by
\beq
\wh g(\tau^1,\tau^2) = \langle \wh e^{-1}(\tau^1),\wh e^{-1}(\tau^2)
\rangle \quad \forall \; \tau^1,\tau^2 \in T_A \pi^{-1}(U(A_0)).
\eeq

Corresponding to the Langevin equation (4.25) this metric 
appears when considering the associated Fokker--Planck operator $L$. We  
rewrite it by a simple manipulation so that it becomes similar to a
Fokker--Planck operator for a stochastic process on a manifold described 
by the metric $\wh g$.
\beqa
L &=& \frac{\delta }{\delta A}\left[  \frac{\delta S}{\delta A} - D_A 
\alpha - 
\frac{\delta^2 A}{\delta \Psi \delta \Psi} (\wt G^{-1} - G^{-1}) +
\frac{\delta}{\delta A} \wh g^{-1} \right] 
\no \\[6pt]
&=& \frac{\delta}{\delta A}\left\{ \wh g^{-1}\left[ \frac{\delta 
S}{\delta A} - ({\bf 1} - \wh g ) \frac{\delta S}{\delta A} - 
 \wh g D_A \alpha + \frac{\delta}{\delta A} \right]\right. \no \\[6pt]
&& - \left. \frac{\delta^2 A}{\delta \Psi \delta \Psi} (\wt G^{-1} -
G^{-1}) +
\frac{\delta \wh g^{-1}}{\delta A} \right\}
\eeqa       
Using the gauge invariance of the Yang--Mills action 
$D_A^* \dfrac{\delta S}{\delta A} =0$  and (5.2)-(5.3) we find
\beq
({\bf 1} - \wh g ) \frac{\delta S}{\delta A} - 
 \wh g D_A \alpha = \wh g D_A (\beta \frac{\delta S}{\delta A} - \alpha
)
\eeq
so that instead of the one form  
$\langle D_A \alpha(A),\cdot\rangle_{(1)}$ corresponding to the original 
Zwanziger term the modified one form given in (5.5) appears in the 
Fokker-Planck operator. The last two terms in (5.4) arise due to the
rules 
of Ito-stochastic calculus; they will turn out later on to give a 
contribution of the form $D_A\xi$.

At this point we want to draw the attention on 
the appearance of the metric $\wh g$.
Since any of the $\wh g$ (parametrized by the yet not specified $\beta$) 
gives rise to a 
specific connection one has an analogous obstruction as in (3.9) when 
trying to have (5.5) 
as a closed one form. A necessary requirement to 
overcome this 
obstruction is therefore that the corresponding curvature has to vanish.
The question how to find such a metric $\wh g$
is reduced to the question how to find a flat 
connection.

Indeed, we can show now that there exists a flat connection in our 
bundle.
The gauge fixing surface $\Sigma$ gives rise to a natural notion of 
horizontal vector spaces in the bundle $\pi^{-1}(U(A_0)) \ra U(A_0)$, 
by declaring all
those vectors $\tau \in T_A \pi^{-1}(U(A_0))$ in the tangent space in 
$A \in \pi^{-1}(U(A_0)) $ to be 
horizontal, which can be written in the form $\tau = \mbox{ad}(g^{-1})
\zeta_B$, where $A = B^g$ and $\zeta_B \in T_B \Sigma$ is a tangent
vector
of $\Sigma$ in point $B$. Let us denote the corresponding subbundle by 
$\wt \Ha$. It is evident by inspection that the corresponding connection
one
form $\wt \gamma$ is given by the following expression
\beq
\wt \gamma = \mbox{ad}(g^{-1}) \F_B^{-1} D_{A_0}^* \mbox{ad}(g), \qquad
A = B^g.
\eeq
This connection is the pull-back of the Maurer--Cartan form
$\theta = \mbox{ad}(g^{-1})R_g$ on the gauge group via the local
trivialization $\chi$
of the bundle $\pi^{-1}(U(A_0)) \ra U(A_0)$. 
The corresponding curvature vanishes, in
other terms expressing the fact that the horizontal subbundle
$\wt \Ha$ is isomorphic to the tangent bundle $T\Sigma$ and
hence integrable. The projector onto the horizontal subbundle is given
by
\beq
\wt {\bf P} = {\bf 1} - D_A \wt \gamma.
\eeq

It has to be mentioned that the connection $\wt \gamma$ cannot be 
extended to a globally defined flat connection on the whole bundle $\A
\ra \M$ 
due to its 
nontriviality. 

Now 
we shall fix the new metric $\wh g$ in such a way that the already
introduced connection $\wt \gamma$ is exactly the induced connection
imposed by itself. In other words this means that the horizontal
subbundle $\wt \Ha$ should be orthogonal to the gauge orbits
with respect to $\wh g$. In particular the gauge fixing surface is then
orthogonal to the gauge orbits. Hence $\wh g$ has to be chosen such that
\beq
\wh g(\wt {\bf P}(\tau),D_A \xi) = \langle \wh e^{-1}
(\wt {\bf P}(\tau)),\wh e^{-1}(D_A \xi)\rangle _{(1)} = 0
\eeq
$\forall \; \xi \in Lie\G$ and 
$\forall \; \tau \in T_A \pi^{-1}(U(A_0))$. 
Using that
\beq
\wh e^{-1} D_A = D_A({\bf 1} + \beta D_A)^{-1}
\eeq
one has to conclude that $\wh e^{-1} \wt {\bf P}$ must be horizontal
with respect to the connection $\gamma$. Hence $\beta$ has to satisfy
\beq
({\bf 1} - {\bf P}) \wh e^{-1} \wt {\bf P} = 0.
\eeq
Using that ${\bf P} \cdot \wt {\bf P} = {\bf P}$ we finally obtain
\beq
\beta {\bf P} = \gamma - \wt \gamma.
\eeq
Notice that $\beta$ is only fixed on the horizontal bundle with respect
to $\gamma$. In vertical direction, however, $\beta$ has only to satisfy
that ${\bf 1} + \beta D_A \neq 0$ in order to guarantee the 
existence of $\wh e^{-1}$. 
The solution for $\beta$ thus obtains as
\beq
\beta = c \gamma - \wt \gamma, \qquad c \neq 0.
\eeq
where $c$ is a non singular map from $Lie~\G$ to $Lie~\G$.
Obviously there is left a freedom for the choice of 
$\beta$ along the gauge group. This, however,
can be proven to be irrelevant in the 
derivation of the equilibrium distribution of the Fokker--Planck 
equation, out of which the path integral density is constructed.
Choosing for $c$ the identity operator the following appealing
expressions for the vielbein $\wh e^{-1}$
\beq
\wh e^{-1} = {\bf 1} - D_A(\gamma - \wt \gamma) =
{\bf P} + ({\bf 1} - \wt {\bf P})
\eeq
as well as for the metric $\wh g$
\beq
\wh g = \langle {\bf P(\tau_1)},{\bf P(\tau_2)}\rangle _{(1)} + 
\langle {(\bf 1} - \wt {\bf P})(\tau_1),({\bf 1} - \wt {\bf 
P})(\tau_2)\rangle _{(1)}
\eeq
are easily derived. 
Notice that $\langle {\bf P(\cdot)},{\bf P(\cdot)}\rangle _{(1)}$
induces a 
metric on the space $\M$.

Similarly as in the case of the helix model there does not exist a 
coordinate transformation $\phi$ such that the Jacobian gives rise
to the vielbein $\wh e^{-1}$. 
In order to prove this fact let us assume the contrary, i.e.
\beq
\wh e^{-1}(\tau) = \phi_* \tau = T \phi(\tau_{\phi^{-1}})
\eeq
for $\tau \in T_A \pi^{-1}(U(A_0))$. But then we get for all vector
fields $\tau^1,\tau^2$ on $\pi^{-1}(U(A_0))$
\beq
[\wh e^{-1}(\tau^1),\wh e^{-1}(\tau^2)] - \wh e^{-1}([\tau^1,\tau^2]) =
[\phi_* \tau^1,\phi_* \tau^2] - \phi_*[\tau^1,\tau^2] = 0
\eeq
using the properties of the push-forward $\phi_*$. On the other hand,
using (5.13) we find for $\tau^1,\tau^2$ being horizontal with respect
to $\wt \gamma$ that the above difference of commutators gives
\beq
[\wh e^{-1}(\tau^1),\wh e^{-1}(\tau^2)] - \wh e^{-1}([\tau^1,\tau^2]) =
[{\bf P}(\tau^1),{\bf P}(\tau^2)] - {\bf P}[\tau^1,\tau^2].
\eeq
That this expression is not vanishing can explicitly be shown by
applying the connection $\gamma$ on the left hand side of (5.17),
yielding
\beq
\gamma
([\wh e^{-1}(\tau^1),\wh e^{-1}(\tau^2)] - \wh e^{-1}([\tau^1,\tau^2]))
=
- \Omega(\tau^1,\tau^2)
\eeq
hence proving that (5.15) cannot be true.

In the adapted coordinates the orthogonality condition of the gauge
fixing
surface and the gauge orbit with respect to the metric $\wh g$ is
transformed into simply
\beq
(\wt G^{-1})^{\Sigma \G} = (\wt G^{-1})^{\G \Sigma} = 0.
\eeq
This condition is fulfilled provided $\beta$ is chosen as above in
(5.12).
Note that for the choice $c=1$ we obtain
\beq
\wt E^\G = D_A^*, \qquad
(\wt G^{-1})^{\G\G} =  \Delta_A .
\eeq

\section{The path integral density as equilibrium distribution}

This section is devoted to the derivation of the Fokker-Planck 
equilibrium distribution which -according to the general principles 
of the stochastic quantization scheme- will be identified with the path 
integral density for the Yang--Mills field.

We previously have already worked out in (4.23) the Langevin equation
for the 
$B$-field. Now we derive from the general Langevin equation (4.15),
inserting the special value (5.20), the corresponding $g$-field equation 
\beqa
dg &=& \left[ -(G^{-1})^{\G \Sigma} \frac{\delta S}{\delta B} + 
\frac{1}{\sqrt{\det G}} \frac{\delta((G^{-1})^{\G \G}\sqrt{\det G})}
{\delta g} + \frac{1}{\sqrt{\det G}} \frac{\delta((G^{-1})^{\G
\Sigma}\sqrt{\det G})}
{\delta B}\right. \no \\[9pt] 
&& \left. + R_g^{-1} \mbox{ad}(g)\alpha\right]ds 
 + \wt E^\G dW.
\eeqa
We choose $\alpha$ as
\beqa
\alpha &=& \mbox{ad}(g^{-1}) R_g \left[ - (\wt
G^{-1})^{\G\G}\frac{\delta 
S_{\G}[g]}{\delta g}
+(G^{-1})^{\G \Sigma}
\frac{\delta S}{\delta B} - \frac{1}{\sqrt{\det G}}
\frac{\delta((G^{-1})^{\G\G}\sqrt{\det G})}{\delta g}
 \right.\no \\[6pt]
&& \mbox{}  \left. -
\frac{1}{\sqrt{\det G}} \frac{\delta((G^{-1})^{\G \Sigma}\sqrt{\det G})}
{\delta B} + \frac{1}{\sqrt{\det G}} 
\frac{\delta((\wt G^{-1})^{\G\G}\sqrt{\det G})}{\delta g} \right]
\eeqa
where $S_{\G}[g]$ is an arbitrary damping function with the
property that
\beq
\int_{\bf \G} \D g \sqrt{\det(R_g^* R_g)} \;
e^{-S_{\G}[g]} < \infty.
\eeq
The choice of $\alpha$ is in fact suggestive: the drift term of 
the $g$-field Langevin equation (6.1) 
is totally exchanged by the damping term $- (\wt
G^{-1})^{\G\G}\dfrac{\delta 
S_{\G}[g]}{\delta g}$; in addition a judiciously chosen Ito--term 
$\dfrac{1}{\sqrt{\det G}} 
\dfrac{\delta((\wt G^{-1})^{\G\G}\sqrt{\det G})}{\delta g}$ is added. 
Due to the choice (6.2) and (6.3) $\alpha$ serves as 
integrating factor to obtain the 
well damped Langevin equation
\beq
dg = \left[ - (\wt G^{-1})^{\G\G} \frac{\delta S_{\G}[g]}{\delta g}
+ \frac{1}{\sqrt{\det G}} 
\frac{\delta((\wt G^{-1})^{\G\G} \sqrt{\det G})}{\delta g} \right]ds
+ \wt E^\G dW.
\eeq
For fixed $B$, the above equation describes a stochastic process on the 
gauge group. 

The Langevin equations (4.23) and (6.4) for $B$ and $g$, respectively, 
can be recast into
\beq
d\Psi = \left[- \wt G^{-1} \frac{\delta S_{\rm tot}[\Psi]}{\delta \Psi}
+ \frac{1}{\sqrt{\det G}} 
\frac{\delta(\wt G^{-1} \sqrt{\det G})}{\delta \Psi} \right] ds
+ \wt E dW
\eeq
where
\beq
S_{\rm tot}[\Psi] = S[B] + S_{\G}[g].
\eeq
The associated Fokker--Planck equation  is derived in a 
straightforward manner
\beq
\frac{\partial \rho[\Psi,s]}{\partial s} = L[\Psi] \rho[\Psi,s],
\eeq
where now the Fokker-Planck operator $L[\Psi]$ is 
appearing in just factorized form
\beq
L[\Psi] = \frac{\delta}{\delta \Psi}
\wt G^{-1} \left[ \frac{\delta S_{\rm tot}[\Psi]}{\delta \Psi}
- \frac{1}{\sqrt{\det G}} 
\frac{\delta(\sqrt{\det G})}{\delta \Psi}
+ \frac{\delta}{\delta \Psi}
 \right].
\eeq
Due to the positivity of $\wt G$ the fluctuation dissipation theorem 
applies and the equilibrium Fokker--Planck distribution 
$\rho^{\rm eq}[\Psi]$ 
obtains by direct inspection as
\beqa
\rho^{\rm eq}[\Psi] &=&
\frac{\sqrt{\det G} e^{- S_{\rm tot}[\Psi]}} {\int_{\bf \Sigma \times
\G} 
\D B \D g
\sqrt{\det G} e^{- S_{\rm tot}}} \no \\[12pt]
&=& \frac{\det \F_B e^{-S[B]} \sqrt{\det R_g^* R_g}
e^{-S_{\G}[g]}} {\int_{\bf \Sigma} \D B \det \F_B e^{-S[B]}
\int_{\bf \G} \D g \sqrt{\det R_g^* R_g} e^{-S_{\G}[g]}}.
\eeqa

This result is completely equivalent to the Faddeev--Popov prescription 
for Yang--Mills theory. The additional $finite$ contributions of the
gauge 
degrees of freedom always cancel out when evaluated on gauge invariant 
observables.  

On the bundle $\pi^{-1}(U(A_0)) \ra U(A_0)$ in the original 
coordinates the Langevin 
equation takes the simple form 
\beq
dA = \left[- \wh g^{-1} \frac{\delta S_{\rm tot}}{\delta A} +
\frac{\delta \wh g^{-1}}{\delta A}  \right] ds + \wh e dW
\eeq
where the total action reads $S_{\rm tot}[A] = S[A] + S_{\G}[\omega 
(A)]$. The dependence of the gauge fixing surface occurs through the
form 
of $\omega (A)$ as defined in (4.4). 
The Fokker--Planck operator in the original coordinates is given by
\beq
L[A] = \frac{\delta}{\delta A}
\wh g^{-1} \left[ \frac{\delta S_{\rm tot}[A]}{\delta A}
+ \frac{\delta}{\delta A}
 \right].
\eeq
With the same argument as above we obtain as a new result
that the equilibrium distribution for the original variables $A$
is given by
\beq
\rho^{\rm eq}[A] = \frac{e^{-S_{\rm tot}[A]}}{\int_{\bf 
\pi^{-1}(U(A_0))} \D A e^{-S_{\rm tot}}}.
\eeq
\section{Outlook}
In this paper we proposed a new stochastic gauge fixing procedure for 
Yang--Mills theory. We were led by the paradigm that instead of the
stochastic 
process itself the expectation values of gauge invariant 
variables should be the right focus. We succeeded in modifying the 
original Parisi--Wu as well as 
Zwanziger's approach such that the Faddeev--Popov path integral density
could be obtained as the Fokker--Planck equilibrium distribution 
in a geometrically transparent way. Distinguished by 
its concept it is the forthcoming task to extend the procedure which so 
far has been 
performed only locally to cover the whole space of gauge potentials.

\end{document}